\documentclass[twocolumn]{aastex63}

\usepackage{xfrac}
\usepackage{graphicx}
\usepackage{color}
\usepackage{amsbsy}
\usepackage{mathrsfs}
\usepackage{mathpazo,bm}
\usepackage{amsmath}
\usepackage{url}
\usepackage{hyperref}
\usepackage{enumitem}
\usepackage{natbib}
\usepackage{amssymb}

\newcommand{\beq}{\begin{equation}}
\newcommand{\eeq}{\end{equation}}
\newcommand{\beqn}{\begin{eqnarray}}
\newcommand{\eeqn}{\end{eqnarray}}
\newcommand{\pderiv}[2]{\frac{\partial{#1}}{\partial{#2}}}

\newcommand{\ptderiv}[1]{\frac{\partial{#1}}{\partial{t}}}

\newcommand{\kderiv}[1]{\frac{{\mathcal D} #1}{{\mathcal D}t}}

\newcommand{\St}{\mathrm{St}}
\newcommand{\Sc}{\mathrm{Sc}}

\newcommand{\ttimes}[1]{10^{#1}}
\newcommand{\xtimes}[2]{#1\times{10^{#2}}}

\renewcommand{\v}[1]{{\boldsymbol{#1}}} 

\newcommand{\ksi}{\xi}

\newcommand{\del}{\v{\nabla}}
\newcommand{\grad}{\del}
\newcommand{\Div}{\del\cdot}
\newcommand{\curl}{\del\times}

\newcommand{\hatz}{\hat{\v{z}}}
\newcommand{\haty}{\hat{\v{y}}}
\newcommand{\hatx}{\hat{\v{x}}}

\newcommand{\cv}{c_{_{V}}}
\newcommand{\cp}{c_p}

\newcommand{\Eq}[1]{Eq.~(\ref{#1})}

\newcommand{\eq}[1]{\Eq{#1}}

\newcommand{\Fig}[1]{Fig.~\ref{#1}}
\newcommand{\Figp}[1]{(Fig.~\ref{#1})}
\newcommand{\figp}[1]{\Figp{#1}}

\renewcommand{\table}[1]{Table~\ref{#1}}
\newcommand{\sect}[1]{Sect.~\ref{#1}}

\definecolor{brown}{rgb}{0.42,0.24,0.07}
\definecolor{darkgreen}{rgb}{0.0,0.6,0.00}
\definecolor{purple}{rgb}{0.7,0.0,0.7}
\definecolor{black}{rgb}{0.0,0.0,0.0}

\def\apj{\rm ApJ}
\def\apjl{\rm ApJL}
\def\apjs{\rm ApJS}
\def\aj{\rm AJ}
\def\mnras{\rm MNRAS}
\def\nat{\rm Nature}

\def\pasp{\rm PASP}
\def\aap{\rm A\&A}

\def\apss{\rm APSS}

\def\icarus{\rm Icarus}

\received{Dec 23, 2020}
\accepted{Apr 10, 2021}

\submitjournal{ApJ}

\shorttitle{3D vortex trapping}
\shortauthors{Raettig et al.}

\begin{document}

\title{Pebble trapping in vortices: three-dimensional simulations}

\correspondingauthor{Wladimir Lyra}
\email{wlyra@nmsu.edu}

\author{Natalie Raettig}
\affiliation{Max-Planck-Institut f\"ur Astronomie,  K\"onigstuhl 17,
  69117, Heidelberg, Germany}

\author[0000-0002-3768-7542]{Wladimir Lyra}
\affiliation{New Mexico State University, Department of Astronomy, PO
  Box 30001 MSC 4500, Las Cruces, NM 88001, USA}

\author[0000-0002-8227-5467]{Hubert Klahr}
\affiliation{Max-Planck-Institut f\"ur Astronomie,  K\"onigstuhl 17,
  69117, Heidelberg, Germany}

\begin{abstract}

Disk vortices have been heralded as promising routes for planet formation due to their ability to trap significant amounts of pebbles. While the gas motions and trapping properties of two-dimensional vortices have been studied in enough detail in the literature, 
pebble trapping in three dimensions has received less attention, due
to the higher computational demand. Here we use the {\sc Pencil Code}
to study 3D vortices generated by convective overstability and the
trapping of solids within them. The gas is unstratified whereas the
pebbles settle to the midplane due to vertical gravity. {We find
  that} for pebbles of normalized friction times of $\St = 0.05$ and $\St = 1$, and dust{-}to{-}gas ratio $\varepsilon=0.01$, the vortex column in the midplane is strongly perturbed. Yet, when the initial dust-to-gas ratio is decreased the vortices remain stable and function as efficient {pebble} traps. Streaming instability is triggered even for the lowest dust-to-gas ratio ($\varepsilon_0 = 10^{-4}$) and smallest {pebble} sizes ($\St = 0.05$) we assumed, showing a path for planetesimal formation in vortex cores from even extremely subsolar metallicity. To estimate if the reached overdensities can be held together solely by their own gravity we estimate the Roche density at different radii. Depending on disk model and radial location of the {pebble} clump we do reach concentrations higher than the Roche density. We infer that if self-gravity was included for the {pebbles} then gravitational collapse would likely occur. 

\end{abstract}


\section{Introduction} 
\label{sect:introduction}

Vortices, hydrodynamical flow in closed elliptic streamings, have
long been considered as a possible route for planet formation since
\cite{BargeSommeria95,AdamsWatkins95} and \cite{Tanga+96} independently
suggested that they would be exceptional sites for 
trapping {solids aerodynamically}. The shear in a Keplerian disk means that only anticyclonic
vortices survive; being high pressure regions, these type of vortices will behave as
dust traps \citep{Whipple72,Bracco+99,Chavanis00}, concentrating the marginally coupled pebbles that would
otherwise drift toward the star \citep{Weidenschilling80,KlahrBodenheimer06}. 

Dust trapping in 2D was subsequently studied by \cite{InabaBarge06},
finding that solids rapidly sink toward the center
of the vortices and increase the solid{s}-to-gas ratio by at least an 
order of magnitude. \cite{Lyra+08b,Lyra+09a,Lyra+09b} {performed} global models, also
in 2D, and including selfgravity, finding that the concentration of
solids collapsed to objects of {lunar and} Mars-mass, that subsequently undergo
further collisions {and pebble accretion}, building up Earth-mass planets.

Because these results were obtained in 2D, where the turbulent
cascade is inverse, the question remained whether a vortex could be
formed and sustained in 3D, where the vortices spontaneously
develop internal turbulence \citep{LesurPapaloizou09}, leading to
direct cascade and vortex destruction. \cite{Lyra09} presented
evidence that vortices would be sustained in 
3D via the Rossby wave instability \citep[RWI,][]{Lovelace+99} in special
locations, namely the boundary between viscosity transitions or the
gaps opened by a planet. Sustenance of Rossby vortices in 3D was subsequently
demonstrated by
\cite{Meheut+10,Meheut+12a,Meheut+12b,Meheut+12c,Lin12a,Lin12b,LyraMacLow12,Lin13}
{and} \cite{Lyra+15}.
Beyond the RWI, \cite{LesurPapaloizou10} {and} \cite{LyraKlahr11} also showed
that in non-isothermal disks, a mechanism for vortex formation exists
in the bulk of the gas, a baroclinic instability, first thought to be
nonlinear, but then found to be linear, and renamed {C}onvective
{O}verstability \citep[{COV,}][]{KlahrHubbard14,Lyra14}. Other non-magnetized
hydrodynamical instabilities in the disk Ohmic zone, the zombie vortex
instability \citep{Marcus+15,Marcus+16,Barranco+18} and the vertical shear
instability \citep{Nelson+13, StollKley14,Richard+16,MangerKlahr18,Flock+20,Manger+20,PfeilKlahr20} also show the
formation of large scale vortices in 3D \citep[see][for a review]{LyraUmurhan19}. 

While these results {from the last decade of research on gas
  dynamics in protoplanetary disks} decisively settled the question of vortex
formation and sustenance in 3D, two-dimensional simulations with dust
cast doubt on the survival of dusty vortices once the dust feedback
becomes too strong {and disrupts the elliptic streamlines} 
\citep{Fu+14,Raettig+15,Surville+16,Miranda+17}. This is a phenomenon that had already been noted by
\cite{InabaBarge06}, that the lifetime of vortices was lowered
as the dust accumulation increased. This has lead to the 
suggestion that the asymmetries seen in ALMA dust continuum
observations of transition disks \citep{vanderMarel+13,vanderMarel+20}
are not vortices, despite the fact that the observational results are 
generally consistent with the analytical predictions for
vortex-trapped dust in steady-state between drag and diffusion 
\citep{KlahrHenning97,LyraLin13,Sierra+17,Casassus+19}. The question
therefore remains whether dusty vortices survive in 3D. 

{To answer this question, we conduct three-dimensional unstratified
gas simulation of the convective overstability, as in 
\cite{LesurPapaloizou10,LyraKlahr11} {and} \cite{Lyra14}, but with pebbles that are
allowed to sediment due to stellar gravity.} 
As we will see, {the dust vertical} stratification dramatically changes the result of  two-dimensional
models, since
pebbles {sediment with a different scale height than the gas},
undergo vertical motion and are subject to vertical
stirring. Another important factor is how the pebbles affect the
vortex structure, which we already saw in our 2D simulations, most
strongly for {loosely coupled grains} \citep{Raettig+15}. In 3D with sedimentation, three different outcomes are possible:

\begin{enumerate}

\item The vortex column is stable and pebbles accumulate inside the column.

\item The pebbles disrupt the vortex column in the layer they reside around
  the midplane, but do not affect the vortex column above and below
  the pebble layer. The question is then how the remaining vortex
  column will affect the pebbles concentration. 

\item The entire vortex column is destroyed by the pebbles. Pebbles
  settle to the midplane with no significant difference with respect
  to non-turbulent runs. The baroclinic vortex cannot be
  reestablished. 

\end{enumerate}

As shown in \cite{Lyra+18}, the second option is the dominant
outcome. Here we detail and expand this result; the phenomenology
depends strongly on {the} pebble radius {and the dust-to-gas
ratio}. First we will introduce
in \sect{sect:physicalbackground} the new physics that is needed for three-dimensional estimates. The numerical
setup is presented in \sect{sect:numericalsetup}. In \sect{sect:results} we will
analyze {the evolution of pebbles in the} simulations and, finally, discuss our results.

\begin{figure*}
  \begin{center}
    \resizebox{.7\textwidth}{!}{\includegraphics{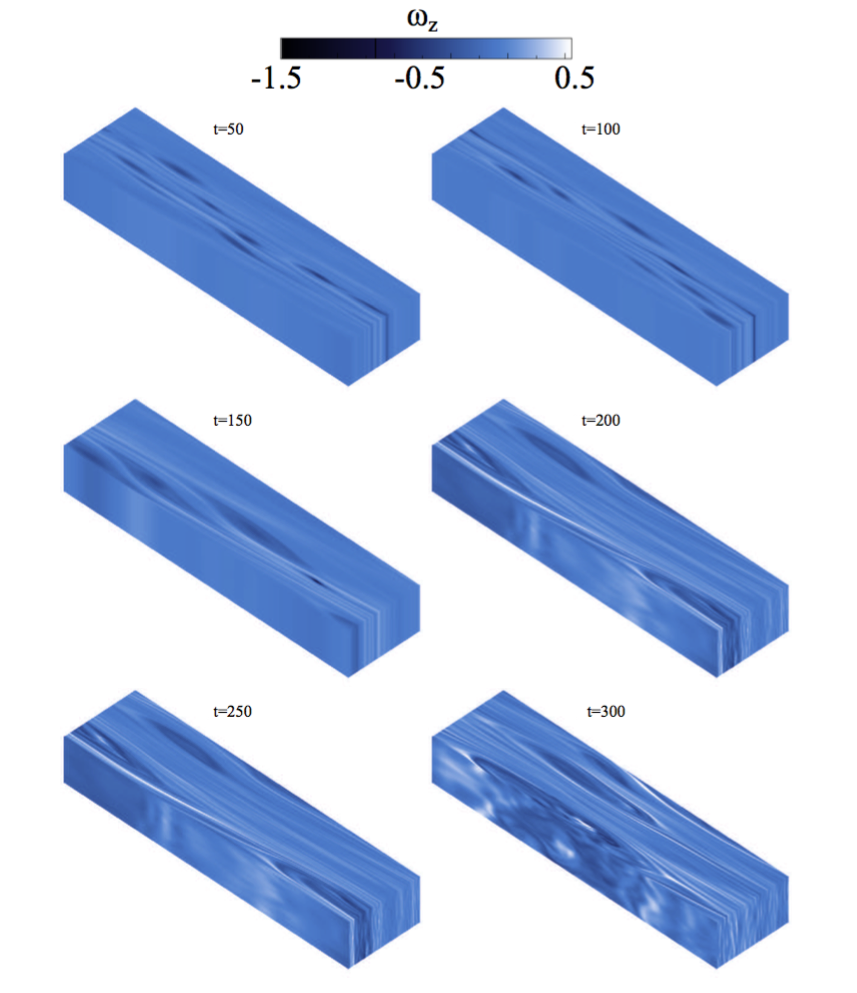}}
\end{center}
\caption{Vertical vorticity of unstratified gas in a three-dimensional
  box at different times. The convective overstability amplifies seed
  noise into growing vortices, that intensity and merge viscously,
  saturating at radial length similar to the pressure scale height.
}
\label{fig:fig2}
\end{figure*}

\begin{figure*}
  \begin{center}
    \resizebox{.7\textwidth}{!}{\includegraphics{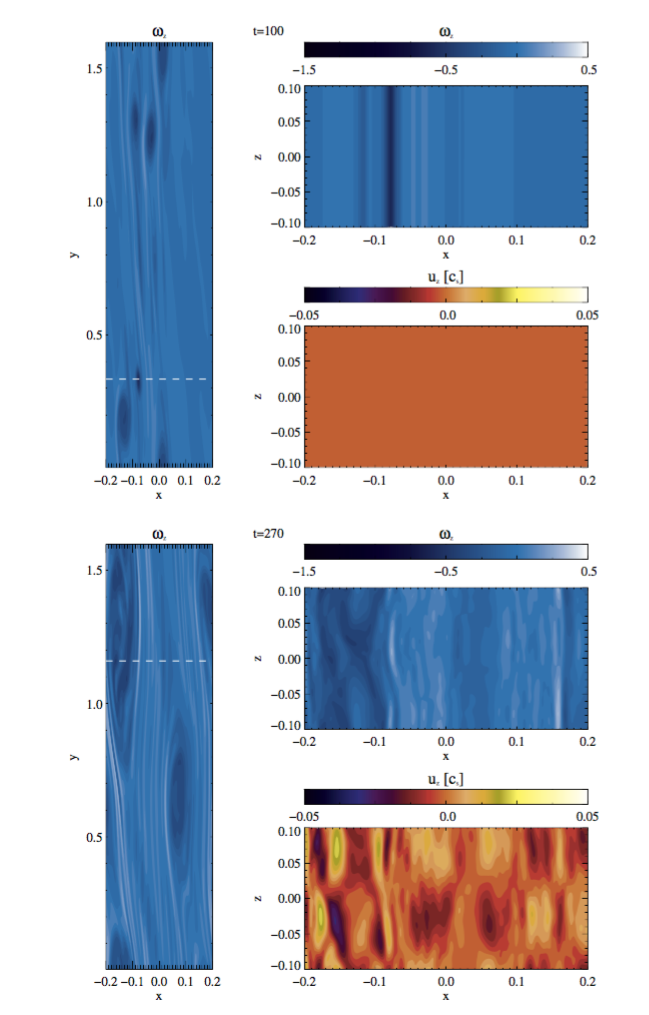}}
\end{center}
\caption{The vertical component of the vorticity ($\omega_z$) in the
  midplane (left panels) and a vertical cut through the disk at the
  azimuthal location of maximum negative vorticity (indicated by the
  dashed line). Also shown is the vertical velocity component ($u_z$). The
  upper plots are shown at 100 local orbits (top panels). At this time,
  there is no structure in the vertical direction: the flow is
  essentially two-dimensional. After 275 local orbits (bottom panels)
  there is now turbulent motion inside the vortex, which is also
  visible in the vertical velocity. This is an indication of the
  growth of elliptical instability in the
  vortex cores.}
\label{fig:fig3}
\end{figure*}

\section{Model Equations}
\label{sect:physicalbackground}

We model a local patch of the disk following the shearing box approximation. The gas is rendered
unstable to the COV by introducing
baroclinicity in the box, in the form of a linearized pressure gradient. The equations of motion are 

\begin{eqnarray}
%
%
  \kderiv{\rho_{{g}}} &=& -\rho_{{g}}\Div\v{u} \label{eq:continuity}\\
%
%
  \kderiv{\v{u}} &=&-\frac{1}{\rho_{{g}}}\grad{p} -2\varOmega\left(\hatz\times\v{u}\right) + \frac{3}{2}\varOmega u_x\haty +\frac{\rho_d}{\rho_{{g}}}\frac{\left(\v{v}-\v{u}\right)}{\tau}  \nonumber\\
&& + f(\beta) \left(\frac{1}{\rho_{{g}}} - \frac{1}{\rho_{{g}0}}\right)\hatx\label{eq:navier-stokes}\\
%
%
      \kderiv{s} &=&\frac{1}{\rho_{{g}}
                     T}\left[\Div\left(K\grad{T}\right) -
                     \rho_{{g}}
                     \cv\frac{\left(T-T_0\right)}{\tau_c}
                     +f(\beta)\frac{u_x}{(\gamma-1)} \right]
                     \nonumber\\
&&\label{eq:entropyeq}
\end{eqnarray} 

In the equations above, {$\rho_g$ is the gas density} $\v{u}$ is
the velocity, {${\varOmega}$ is the Keplerian frequency, $\rho_d$ is
the dust density, $\v{v}$ is the dust velocity, $\tau$ is the friction
time between gas and dust}, $s$ is the entropy, $T$ is the
temperature, $K$ is the radiative conductivity, {$\cv$ is the heat
capacity at constant volume, $\gamma=c_p/\cv$ is the adiabatic index,
$c_p$ is the heat capacity at constant pressure, $T_0$ is the gas equilibrium temperature,
and $\tau_c$ is the cooling time}. The operator 

\begin{equation}
  \kderiv{} \equiv \pderiv{}{t} {+\v{u}\cdot\del} + u_{{k}}\pderiv{}{y}
\end{equation}represents the {Lagrangian} derivative of a fluid
parcel. It is the only place where the {(linearized)} Keplerian flow
$\v{u}_{{k}}{=-\sfrac{3}{2}\,\varOmega x\,\hat{y}}$
appears explicitly. The function $f(\beta)$ is

\beq
f(\beta) \equiv \frac{\beta p_0 \varOmega}{c_{s0}}
\eeq

\noindent where the parameter $\beta \equiv  h\xi$, {(with $h=c_s/u_k$ the
disk aspect ratio, and $c_s$ the sound speed)}, is related to the power
law of the pressure gradient

\beq
\bar{p}(r) \equiv p_0 (r/r_0)^{-\xi} ,
\eeq

\noindent where $r$ is the cylindrical radius and $r_0$ is a reference radius. The overbar indicates that this quantity is time-independent.
The total pressure is $p_{\rm tot}$=$\bar{p}+p$, where $p$ is the
local fluctuation. Linearizing {this pressure gradient} contributes an extra term in the
momentum equation and in the energy equation, respectively. The
derivation of this extra term and the limitations of the approach are
detailed in \cite{LyraKlahr11}. {All the symbols used in this work are
listed in \table{table:symbols}.} 

The advection is made Galilean-invariant by means of the SAFI algorithm \citep{Johansen+09a}. The simulations 
were done with the {\sc Pencil Code} 
{\footnote{See http://pencil-code.nordita.org}} \citep{BrandenburgDobler02,BrandenburgDobler10,Brandenburg+20}, a collocated finite-difference code which integrates the partial differential equations with sixth-order spatial derivatives, and a third-order Runge–Kutta time integrator. We add explicit sixth-order hyperdiffusion, hyperviscosity,
and hyper heat conductivity to the mass, momentum, and entropy
equations, respectively, explained in
\cite{Lyra+08a,Lyra+09a,Lyra+17}. They are needed because the
high-order scheme of the Pencil Code has little overall numerical
dissipation \citep{McNally+12}. {We use dimensionless code units;
with} all simulations us{ing} $c_p=\varOmega =\rho_{{g}0}=1$, $\gamma=1.4$, $c_{s0}=h=0.1$, and
$\xi=2$. {The models are scale-free and, thus, the translation to
  physical units depends on specific parameter choices, as well as on the distance
  from the box to the star. For densities of the Minimum-Mass Solar
  Nebula, \citep[MMSN,][]{Weidenschilling77,Hayashi81}, $\Sigma_g = 1700
  (r/{\rm AU})^{-1.5} {\rm g\,cm}^{-2}$ and temperature of
  100\,K at 5AU and 30\,K at 50\,AU, the disk is stable
  against gravitational instabillity \citep{Safronov60,Toomre64}, having a Toomre
  $Q$ value of  50 at 5\,AU and 20\, at 50AU.}

\subsection{Pebble evolution}

The pebbles are represented by numerical particles that evolve
Lagrangianly according to 

\beqn
\frac{d\v{x}}{dt} &=& \v{v}\\
\frac{d\v{v}}{dt} &=& -\frac{\left(\v{v}-\v{u}\right)}{\tau}
-2\varOmega \left(\hatz\times\v{v}\right) +
\frac{3}{2}\varOmega v_x\haty -\varOmega^2 z \hatz\nonumber\\
\eeqn

\noindent where {$\v{x}$ is the pebble position and} $\v{v}$ 
is the pebble velocity corrected by the sub-Keplerian parameter

\beq
\eta \equiv \frac{1}{2}h^2 \xi{.}
\eeq

\noindent The extra velocity $\Delta v = \eta u_k$ is added to the {pebbles}, as explained in
\cite{Raettig+15}. This term is needed because the frame of reference
is sub-Keplerian, as ensured by the extra $1/\rho_{{g}0}$ term in
\eq{eq:navier-stokes}. In the reference frame of the box, the
sub-Keplerian gas is stationary, and the {pebbles} flow with 
$\eta u_k$ azimuthal velocity. This is constrated with simulations by
\cite{Johansen+06} where the reference frame is Keplerian, i{.}e{.} the
{pebbles} are stationary and the gas flows with 
$-\eta u_k$ azimuthal velocity. Here we use $\eta=0.01$, which
corresponds to a pressure gradient $\xi=2$ and disk aspect ratio
$h=0.1$. The terminal velocities of the {pebbles} for a
given Stokes number {relative to the Keplerian flow} are

\beqn
v_x &=& -\frac{2\eta u_{{k}}}{\St + \St^{-1}}\\
v_y &=& -\frac{\eta u_{{k}}}{{1+}\St^2}
\eeqn

In the vertical direction, the {pebbles} settle with a constant velocity so that $\partial_t v_z \approx 0$. This leads to

\beq
\pderiv{z}{t} = -\tau \varOmega^2 z 
\eeq

and

\beq
z{(t)} = z_0 e^{-t\tau\varOmega^2}.
\label{eq:zscale}
\eeq

\noindent Here {pebbles} will approach the midplane exponentially but never reach
it. For {pebbles} with $\St \gg 1$ the term $\tau^{-1}v_z$ is negligible and the equation
of motion reduces to

\beq
\pderiv{v_z}{t} = -\varOmega^2 z
\eeq

\noindent where

\beq
z{(t)} = \cos (\varOmega t)
\eeq

\noindent is a solution. Heavy {solids} will oscillate around the midplane, never settling permanently. In a real protoplanetary disk there are additional effects acting on the {solids} that will change this idealized picture.
The sedimentation time scale $t_s$ for small {grains} can be deduced
from \eq{eq:zscale}. After a time $t_s = (\varOmega\,\St)^{-1}$ a
{pebble} will have sedimented by a factor $e$ of its original
height. The {pebble} stirring works as an
effective diffusion, so that the evolution of the {pebble} density can
be described by

\beq
\ptderiv{\rho_d} = D\pderiv{}{z}\left[\rho_g\pderiv{}{z}\left(\frac{\rho_d}{\rho_g}\right)\right] + \pderiv{}{z}\left(\varOmega^2\tau_s\rho_d z\right),
\eeq

\noindent where $D$ is the diffusion constant. For steady-state
\citep{Dubrulle+95,KlahrHenning97,LyraLin13}, the pebble distribution
settles as a Gaussian around the midplane, with pebble scale height

\beq
H_d = H \sqrt{\frac{\delta}{\St+\delta}}
\label{eq:dust-scale-height}
\eeq

\noindent where {{$H \equiv hr_0 \equiv c_s/\varOmega$ is the gas scale
height and}}  $\delta \equiv D/(c_s H)$ is a dimensionless
diffusion coefficient similar to the Shakura-Sunyaev $\alpha$-viscosity
coefficient \citep{ShakuraSunyaev73}. The ratio of $D$ and $\nu$ defines the dimensionless Schmidt number

\beq
\Sc \equiv \frac{\nu}{D} = \frac{\alpha}{\delta}
\eeq

The pebble scale height then is 

\beq
H_d = H \sqrt{\frac{\alpha}{\St\,\Sc + \alpha}}
\eeq

\begin{table*}
\caption{{Symbols used in this work.}}
\label{table:symbols}
 \begin{center}
\begin{tabular}{lllclll}\hline
{Symbol}        &{Definition}                                                           &{Description}                       &&{Symbol}        &{Definition}                     &{Description}                        \\\hline
{$\rho_g$}      &{}                                                                     &{gas density}                       &&{$\xi$}         &{$d\ln p/d\ln r$}                &{pressure gradient}                  \\
{$\v{u}$}       &{}                                                                     &{gas velocity}                      &&{$H$}           &{$c_s/\varOmega$}                &{gas scale height}                   \\
{$p$}           &{$\rho_g c_s^2/\gamma$}                                                &{gas pressure}                      &&{$h$}           &{$H/r_0$}                        &{gas aspect ratio}                   \\
{$s$}           &{\tiny{$\cv[\ln(\sfrac{T}{T_0})-(\gamma-1)\ln(\sfrac{\rho}{\rho_0})]$}}&{gas entropy}                       &&{$\beta$}       &{$h\xi$}                         &{scaled pressure gradient}           \\
{$T$}           &{}                                                                     &{gas temperature}                   &&{$\v{\omega}$}  &{$\curl{\v{u}}$}                 &{gas vorticity}                      \\
{$K$}           &{}                                                                     &{radiative conductivity}            &&{$\eta$}        &{$h^2\xi/2$}                     &{sub-Keplerian parameter}            \\
{$c_s$}         &{$[T\cp(\gamma-1)]^{1/2}$}                                             &{sound speed}                       &&{$\St$}         &{$\varOmega \tau$}               &{Stokes number}                      \\
{$\cv$}         &{}                                                                     &{Specific heat at constant volume}  &&{$t_s$}         &{}                               &{sedimentation time}                 \\
{$\cp$}         &{}                                                                     &{Specific heat at constant pressure}&&{$\delta$}      &{}                               &{dimensionless diffusion coefficient}\\
{$\gamma$}      &{$\cp/\cv$}                                                            &{Adiabatic index}                   &&{$D$}           &{$\delta c_s H$}                 &{diffusion coefficient}              \\
{$r_0$}         &{}                                                                     &{Reference radius}                  &&{$H_d$}         &{$H\sqrt{\delta/(\delta + \St)}$}&{dust scale height}                  \\
{$\varOmega$}   &{}                                                                     &{Keplerian frequency at $r_0$}      &&{$\alpha$}      &{}                               &{dimensionless viscosity coefficient}\\
{$\v{u}_k$}     &{$-\sfrac{3}{2}\,\varOmega x\,\hat{y}$}                                &{Keplerian velocity at $r_0$}       &&{$\nu$}         &{$\alpha c_s H$}                 &{viscosity coefficient}              \\
{$r$}           &{}                                                                     &{cylindrical radius}                &&{$\Sc$}         &{$\nu/D$}                        &{Schmidt number}                     \\
{$x$}           &{$r-r_0$}                                                              &{Cartesian radial coordinate}       &&{$\varepsilon$} &{$\rho_d/\rho_g$}                &{dust-to-gas ratio}                  \\
{$y$}           &{}                                                                     &{Cartesian azimuthal coordinate}    &&{$k$}           &{}                               &{wavenumber}                         \\
{$z$}           &{}                                                                     &{vertical coordinate}               &&{$L$}           &{}                               &{box length}                         \\
{$t$}           &{}                                                                     &{time}                              &&{$\Sigma_g$}    &{$\int\rho_g dz$}                &{gas column density}                 \\
{$\rho_d$}      &{}                                                                     &{dust density}                      &&{$\Sigma_d$}    &{$\int\rho_d dz$}                &{dust column density}                \\
{$\v{v}$}       &{}                                                                     &{dust velocity}                     &&{$\varphi$}     &{$[0,1]$}                        &{Random phase}                       \\
{$\tau$}        &{}                                                                     &{drag time}                         &&{$a_\bullet$}   &{}                               &{pebble radius}                      \\
{$\tau_c$}      &{}                                                                     &{cooling time}                      &&{$\rho_\bullet$}&{}                               &{pebble internal density}            \\
{$\rho_R$}      &{$3\pi M_\odot/(2r^3)$}                                                &{Roche density}                     &&{}              &{}                               &{}                                   \\\hline
\end{tabular}
\end{center}
\end{table*}

\begin{table*}
\caption{Simulation setup and results}
\label{table:table1}
\begin{center}
\begin{tabular}{c c c c c c c c c c c c c}\hline
Run & $\St$ & $\varepsilon_0$ & feedback & ${\xi}$ & $H_d$ & $D$& $\delta$ & $\alpha$ & $\Sc_z$ & $t_{\rm end}$ & $\varepsilon_{\rm max}$ & ${\zeta}$\\
& & & & & \footnotesize{$(\times10^{3})$} & \footnotesize{$(\times10^{7})$}& \footnotesize{$(\times10^{5})$} & \footnotesize{$(\times10^{3})$} & & &  & \\\hline
3DG            & -     &-           &- &2 &-&-&-&-&-&-&-&-\\
3DF05nt     &0.05 &  $10^{-2}$ & yes&0   & 1.20 & 0.72 & 0.72 & 0.1 & 13.87 & 43.4&41.6  &-    \\  
3DNF05      &0.05 &  $10^{-2}$& no&2     & 2.57 & 3.31 & 3.31 & 1.12 & 33.83 & 40.7&3540.0 &-    \\  
3DF05        &0.05 &   $10^{-2}$& yes&2  & 2.23 & 2.49 & 2.49 & 0.66 & 26.7 & 38.2&111.2 & 98.44\\
3DF05E-3  &0.05 &  $10^{-3}$& yes&2 &-&-&-&-&-& 25.3&8.7 &25.67\\
3DF05E-4  &0.05 & $10^{-4}$& yes&2 &-&-&-&-&-& 20.0&2.1 &1.12\\
3DNF1       &1.0   &  $10^{-2}$& no &2 & 0.99  & 9.84 & 9.84 & 0.94 & 9.51 & 22.4&15630.0&-\\
3DF1         &1.0    & $10^{-2}$& yes&2& 0.85  & 7.20 & 7.20 & 0.26 & 3.72 & 22.8&945.4 & 98.73\\
3DF1E-3   &1.0    & $10^{-3}$& yes&2 &-&-&-&-&-& 7.7&227.7 & 26.41\\
3DF1E-4   &1.0    & $10^{-4}$ & yes&2 &-&-&-&-&-& 9.7&34.2 & 14.46\\\hline
\end{tabular}
\end{center}
\end{table*}

\section{Numerical Setup}
\label{sect:numericalsetup}

\subsection{Gas Setup}
\label{sect:gassetup}

Since we do not include vertical gravity for the gas simulations, the
gas setup is the same as in \cite{Raettig+15}, with initial noise in the
density according to 

\beq
\rho_{{g}} = C\rho_{{g}0}  e^{-(x/2\sigma)^2} 
\sum_{i=-k_x}^{k_x}\sum_{j=0}^{k_y} \sin \left[ 2\pi
  \left(i\frac{x}{L_x} +j\frac{y}{L_y} + \varphi_{ij}\right)\right]
\eeq

\noindent {where $L$ is the box size, and the phase $0 < \varphi < 1$ determines the randomness. The
  subscripts underscore that the phase is the same for all grid
  points, only changing with wavenumber. The constant $C$ sets the
  strength of the perturbation. The width $\sigma$ is set to $L_x/5$.} In this setup no vertical modes are excited,
but radial and azimuthal modes are excited in all {vertical} layers of the
box. Additionally, we add Gaussian noise to the velocity.
The physical size of our box is the same in radial ($\pm 2H$) and
azimuthal (${\pm 8}H$) direction as in the 2D runs in
\cite{Raettig+13,Raettig+15}. We model $\pm 1H$ in the vertical
direction. This is a compromise between {an} adequately large physical
domain to capture the elliptical instability inside the vortex cores,
but not high enough resolution to resolve all modes that contribute to
the streaming instability. The vertical resolution is the same as in
the radial direction. We restrict ourselves to an entropy gradient of $\xi
= 2$, since we established in \cite{Raettig+13} that the main
difference between $\xi = 2$ and lower entropy gradients was the
{time to reach saturation}. The size of the vortices and strength of relevant
parameters such as {kinetic} stresses depend only weakly on $\xi$. For
$\xi = 2$ a resolution of $256 \times 256 \times 128$ is appropriate
to resolve the COV {to convergence} \citep{LyraKlahr11,Raettig+13,Lyra14}. 

\subsection{{Pebble} Setup}
\label{sect:particlesetup}

For the {pebbles}, we add the vertical equation of motion, also including linear vertical gravity $g_z = -\varOmega^2 z$.
As in the 2D simulations we first evolve the gas alone for 200 local
orbits before adding the pebbles. In these simulations, the
{pebbles} are randomly distributed in the $x-y$ plane as before,
but {now they} follow a Gaussian profile in the vertical direction

\beq
\rho_d(z) = C_d \ e^{-\frac{z^2}{2{H_{d0}}^2}}
\eeq

\noindent where $\rho_d$ is the dust density, $C_d$ a normalization
constant and ${H_{d0}}$ gives the scale height of the initial
distribution. We set ${H_{d0}} = 0.01$, which gives a well
settled profile, and speeds up the initial relaxation.
Like in the 2D simulations \citep{Raettig+15} we want to have about
$4-5$ super-particles per grid point initially, to mitigate numerical
effects {due to Poisson noise} \citep{Johansen+07}. Because most {pebbles} will settle into the midplane, we only
need to take into account
$\pm$10 grid points around the midplane in this estimate, leading to a
particle count of $7\times 10^6$. 

{Pebble} feedback onto the gas is included for all simulations except
for the control runs. We
restrict oursel{ves} to two {pebble} sizes, $\St = 1$ for loosely coupled
{pebbles} and $\St = 0.05$ for more strongly coupled
{pebbles}. {The Stokes number scales with grain size as }

\beq
\St = \frac{a_\bullet \rho_\bullet}{H \rho_g}
\eeq

\noindent {where $a_\bullet$ is the grain radius and $\rho_\bullet$ its internal
  density. {In the simulations we fix $\St$, not the grain
    radius, so effectively $a_\bullet$ varies dynamically with
    $\rho_g$ and $H$ due to compressibility. However, in practice, the
    vortices are not too compressible, with density variations of 50\%
  at most.} For the MMSN and assuming silicate density ($\rho_\bullet=
3\, {\rm g\,cm}^{-3}$), the corresponding grain sizes for $\St=0.05$ at 1,
5, and 50 AU are 10cm, 1cm, and 0.3 mm. For ice internal density
($\rho_\bullet = 1 \, {\rm g\,cm}^{-3}$), the sizes would be scaled by about 3,
so about 30cm, 3cm, and 1 mm. For $\St=1.0$, the grain radii would be
2m, 20cm, and 0.6cm at 1, 5, and 50 AU for silicate density, and 6m,
60cm, and 2cm for ices at the same distances. For a nebula 5 times
more massive than MMSN (Toomre $Q \sim 10$ at 5\,AU and 5 at 50\,AU, 
marginally stable at these distances) the sizes are scaled up by 5. In
short, the $\St=0.05$ particles represent cm-sized pebbles in the
inner disk and mm-sized in the outer disk, whereas $\St=1.0$ particles
represent m-sized boulders in the inner disk and cm-sized pebbles in
the outer disk. While coagulation/fragmentation equilibrium does not
predict the existence of m-sized boulders \citep{Guttler+10,Zsom+10}, the $\St=1$ particles representing
cm-sized range in the outer disk is a pebble size important for planet
formation and captured by VLA observations \citep{Casassus+19}.}

We expect the setup to develop $\alpha \approx
10^{-3}$ \citep{Lyra14}. If we assume for now that
the Schmidt number for our system is 1{,} then the {pebble} scale
height for $\St=0.05$ is $H_d = 0.01$. Therefore, a {pebble} layer of $2H_d$ around the
midplane should be resolved with 12 grid points. The layer of $\St=1$
will not be resolved. We set the initial dust-to-gas ratio to $\varepsilon_0 = 0.01$. Additionally,
we perform runs with $\varepsilon_0 = \ttimes{-3}$ and $\varepsilon_0
= \ttimes{-4}$. We also perform one simulation with $\xi = 0$ for $\St =
0.05$ for a control run without turbulence. All setups are listed in Table~\ref{table:table1}. Particle block domain decomposition is used for load balance \citep{Johansen+11}. 

\begin{figure}
  \begin{center}
    \resizebox{\columnwidth}{!}{\includegraphics{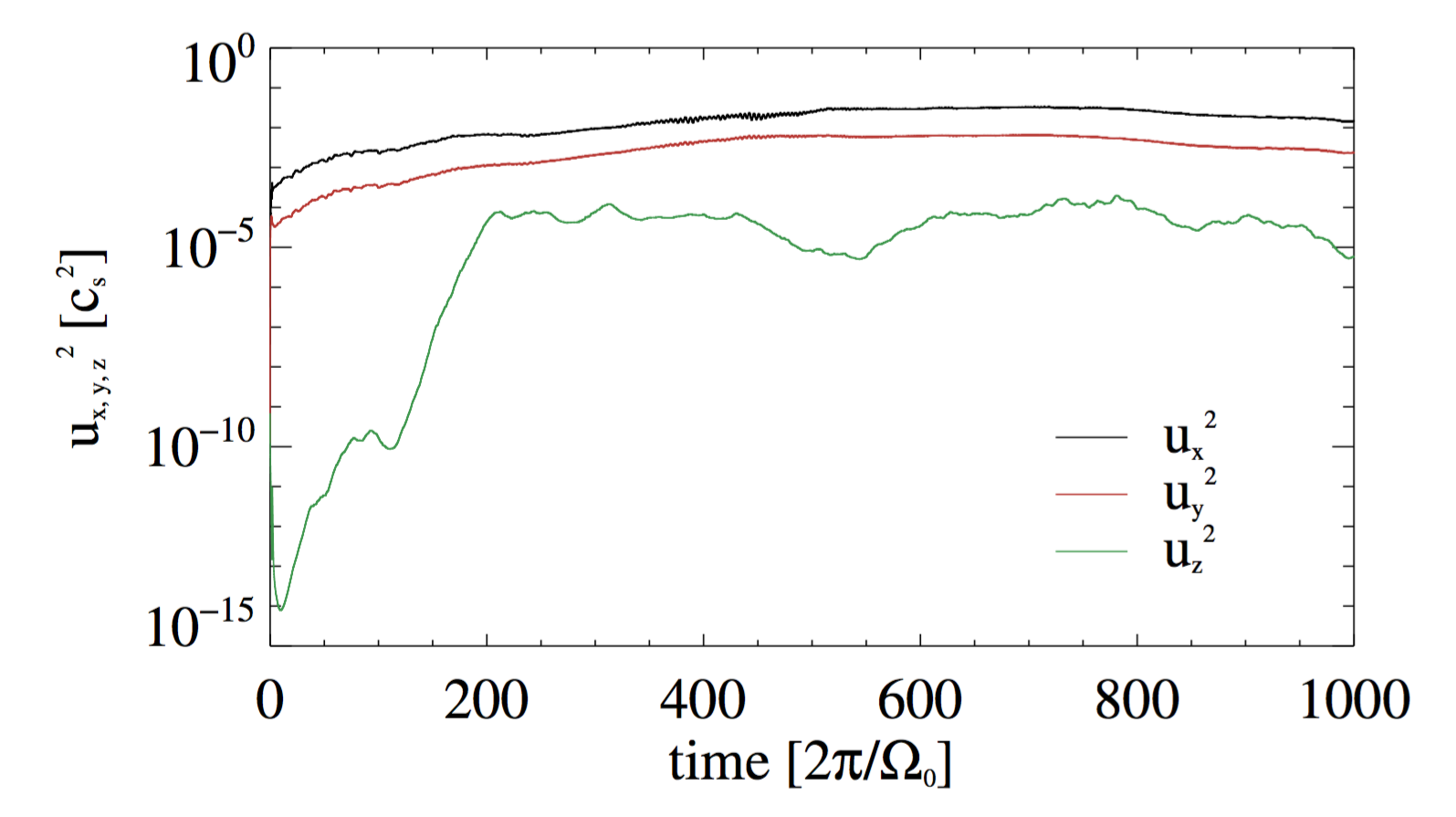}}
\end{center}
\caption{Gas velocity components averaged over the entire box. As the
  elliptical instability sets in, the vertical velocity component
  (green line) increases steeply, {while} stay{ing} well below the azimuthal and radial components.
}
\label{fig:fig4}
\end{figure}

\section{Results}
\label{sect:results}

\begin{figure}
  \begin{center}
    \resizebox{\columnwidth}{!}{\includegraphics{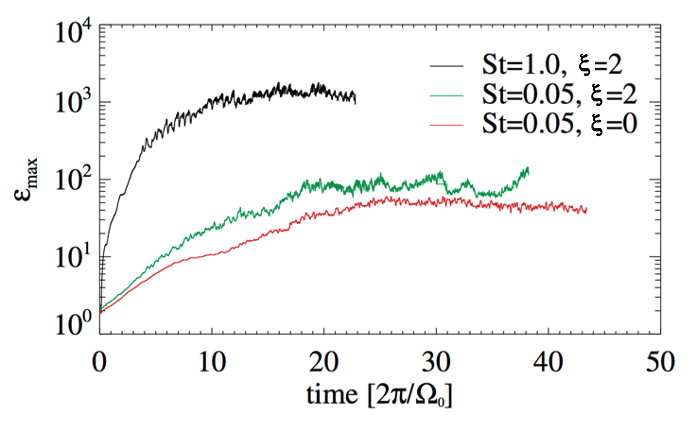}}
\end{center}
\caption{Maximum dust-to-gas ratio for simulations with $\St = 1$ (black line) and $\St = 0.05$ (green and red line) {pebbles}. $\xi = 0$ indicates that no baroclinic feedback was included in the simulation. Although the high {pebble} concentration in the simulations with baroclinic effects disrupts the vortex in the midplane, where {pebbles} are located, the maximum dust-to-gas ratio is higher than without baroclinic effects.
}
\label{fig:fig7}
\end{figure}

\begin{figure*}
  \begin{center}
    \resizebox{.8\textwidth}{!}{\includegraphics{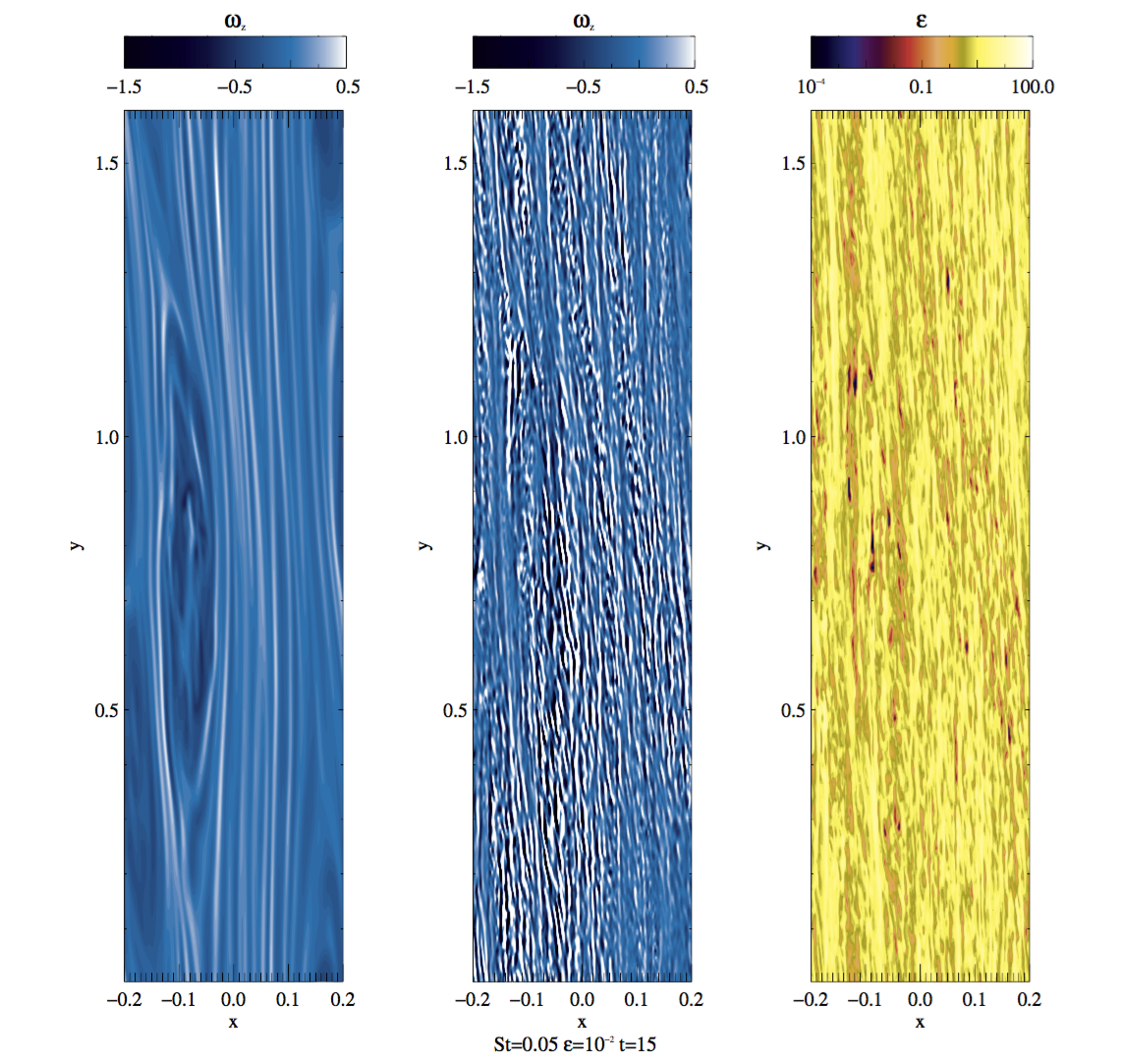}}
\end{center}
\caption{Vertical vorticity component $\omega_z$ at the top of the box (left) and in the midplane (middle) and dust-to-gas ratio $\varepsilon$ (right) in the midplane for $\St = 0.05$ {pebbles} and $\varepsilon_0 = 10^{-2}$. The vortex in the midplane is not apparent, and the {pebbles} are spread out. Since there was an initial {pebble} accumulation inside the vortex before it was disrupted, there is a residual accumulation at the original vortex position. The maximum dust-to-gas-ratio in this snapshot is $\varepsilon = 31.7$.
}
\label{fig:fig8}
\end{figure*}

\begin{figure*}
  \begin{center}
    \resizebox{.8\textwidth}{!}{\includegraphics{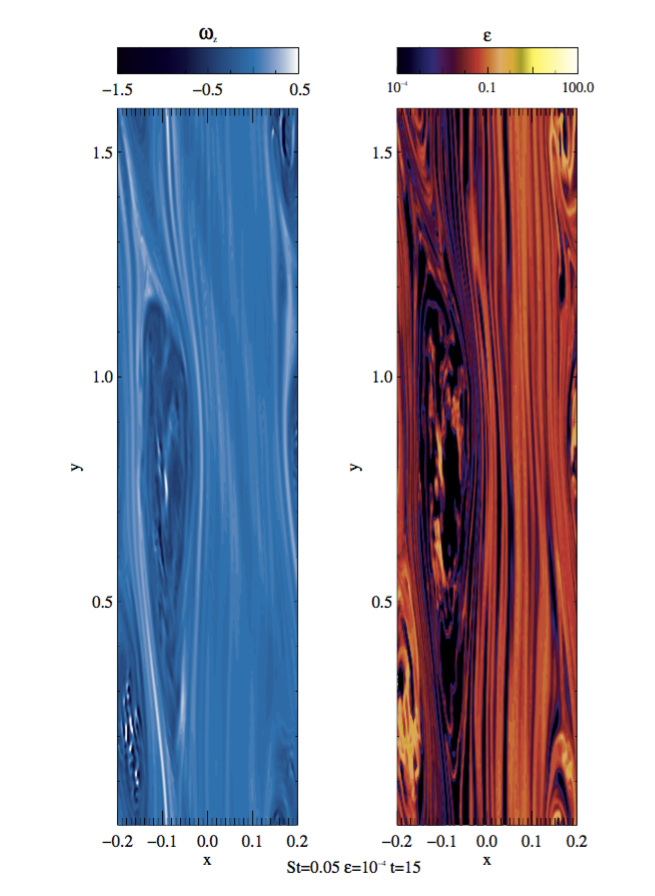}}
\end{center}
\caption{Vertical vorticity component $\omega_z$ and dust-to-gas ratio $\varepsilon$ in the midplane for $\St = 0.05$ {pebbles} and $\varepsilon_0 = 10^{-4}$. In contrast to simulations with $\varepsilon_0 = 10^{-2}$ the vortices in the midplane are not disrupted and the {pebbles} accumulate inside of the two vortices. The maximum dust-to-gas-ratio reached in this snapshot is $\varepsilon = 2.2$ which is one order of magnitude lower than for $\varepsilon_0 = 10^{-2}$.}
\label{fig:fig9}
\end{figure*}

The evolution of the vortex in a 3D unstratified box is very similar
to the two-dimensional case. First, a number of small vortices
emerge. These vortices are then amplified by the convective
overstability mechanism
\citep{LesurPapaloizou10,LyraKlahr11,KlahrHubbard14,Lyra14} and also
merge. \Fig{fig:fig2} shows snapshots of the vorticity in the fiducial
model. At some point the vertical gas motion sets in and generates
turbulent features in the vorticity structure (compare top and bottom
panel of \Fig{fig:fig3}). After about 100 local orbits there is a
steep increase in vertical gas velocity, due to elliptical instability \citep{Kerswell02,LesurPapaloizou09,LesurPapaloizou10,LyraKlahr11,Lyra13}. Yet it remains $2-3$ orders
of magnitude lower than the radial and azimuthal velocity components \figp{fig:fig4}.

\begin{figure*}
  \begin{center}
    \resizebox{.8\textwidth}{!}{\includegraphics{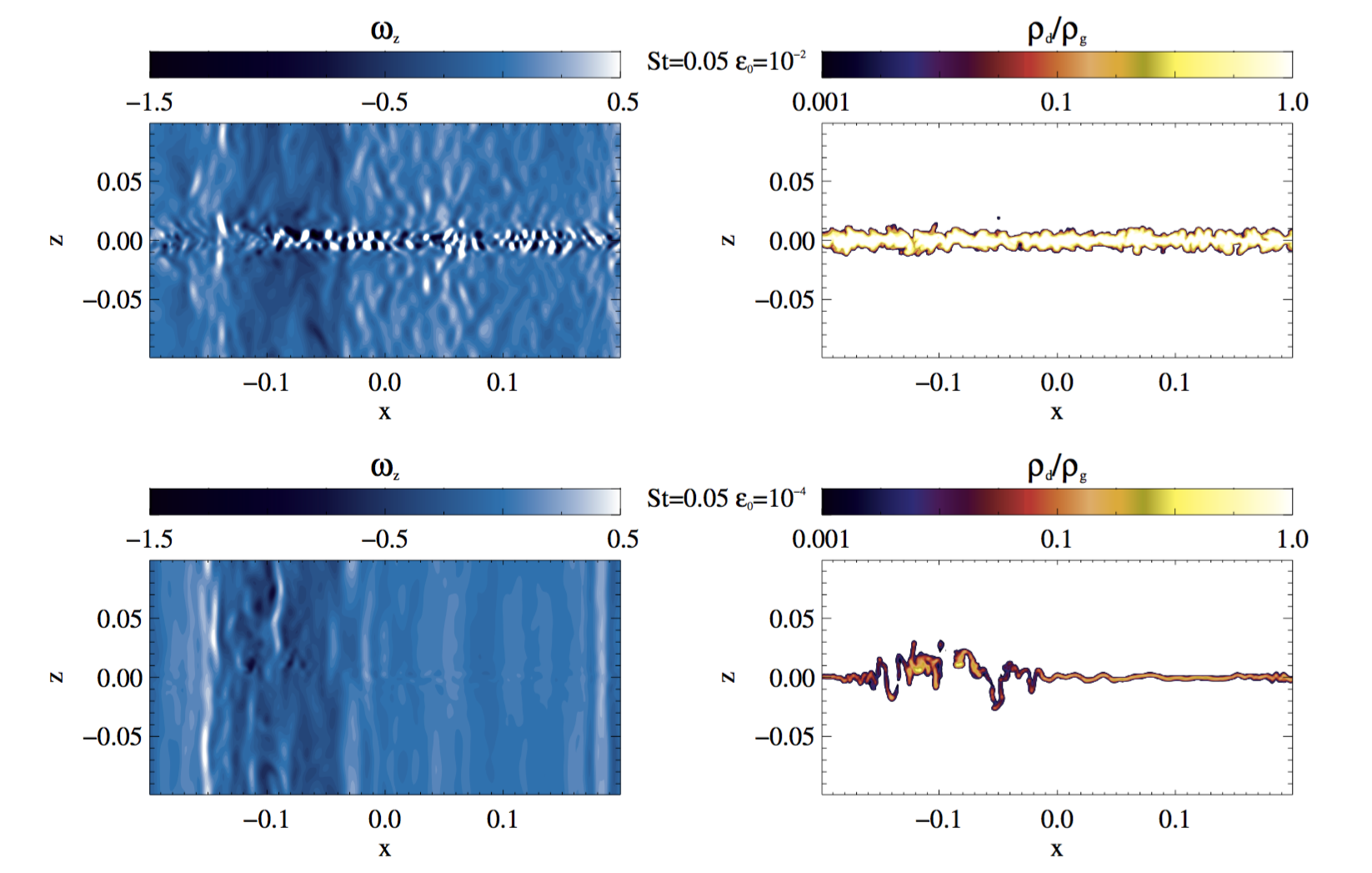}}
\end{center}
\caption{Vertical cut through the vortex after 15 local orbits for $\St = 0.05$ {pebbles} and $\varepsilon_0 = 10^{-2}$ (top) and $\varepsilon_0 = 10^{-4}$ (bottom). Shown are the vertical component of vorticity $\omega_z$ (left) and the dust-to-gas ratio $\rho_d/\rho_g$ (right). For $\varepsilon_0 = 10^{-2}$ the vortex located around $x \approx -0.1H$ is strongly perturbed around the midplane. For $\varepsilon_0=10^{-4}$ the vortex column remains roughly undisturbed.
}
\label{fig:fig10}
\end{figure*}

\begin{figure}
  \begin{center}
    \resizebox{\columnwidth}{!}{\includegraphics{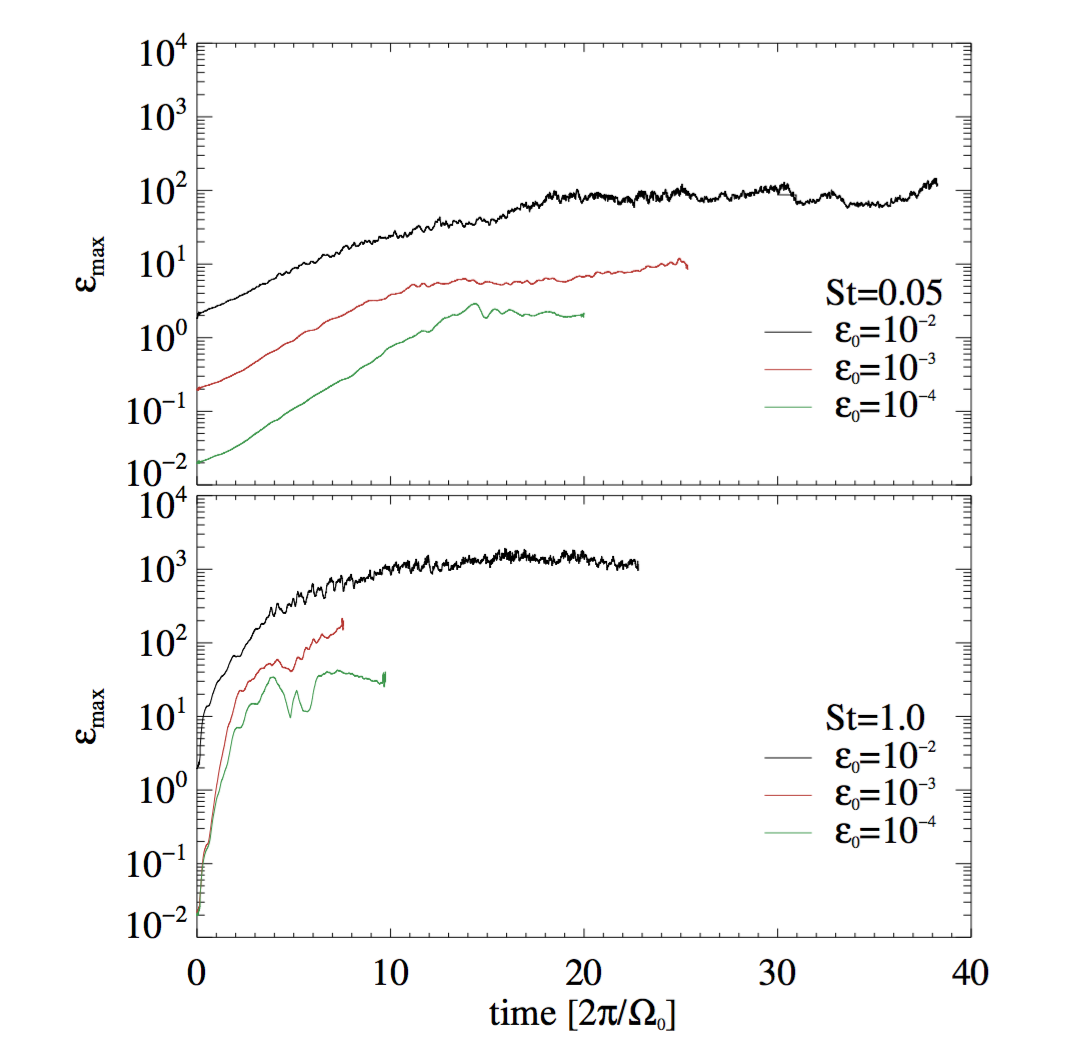}}
\end{center}
\caption{
Maximum dust-to-gas ratio for simulations with $\St = 0.05$ (to panel)
and $\St = 1$ (bottom panel) {pebbles,} and different initial
dust-to-gas ratios $\varepsilon_0$. The concentration
$\rho_d/\rho_{d0}$ seem {to be} only weakly dependent of $\varepsilon_0$.
}
\label{fig:fig11}
\end{figure}

\begin{figure*}
  \begin{center}
    \resizebox{.8\textwidth}{!}{\includegraphics{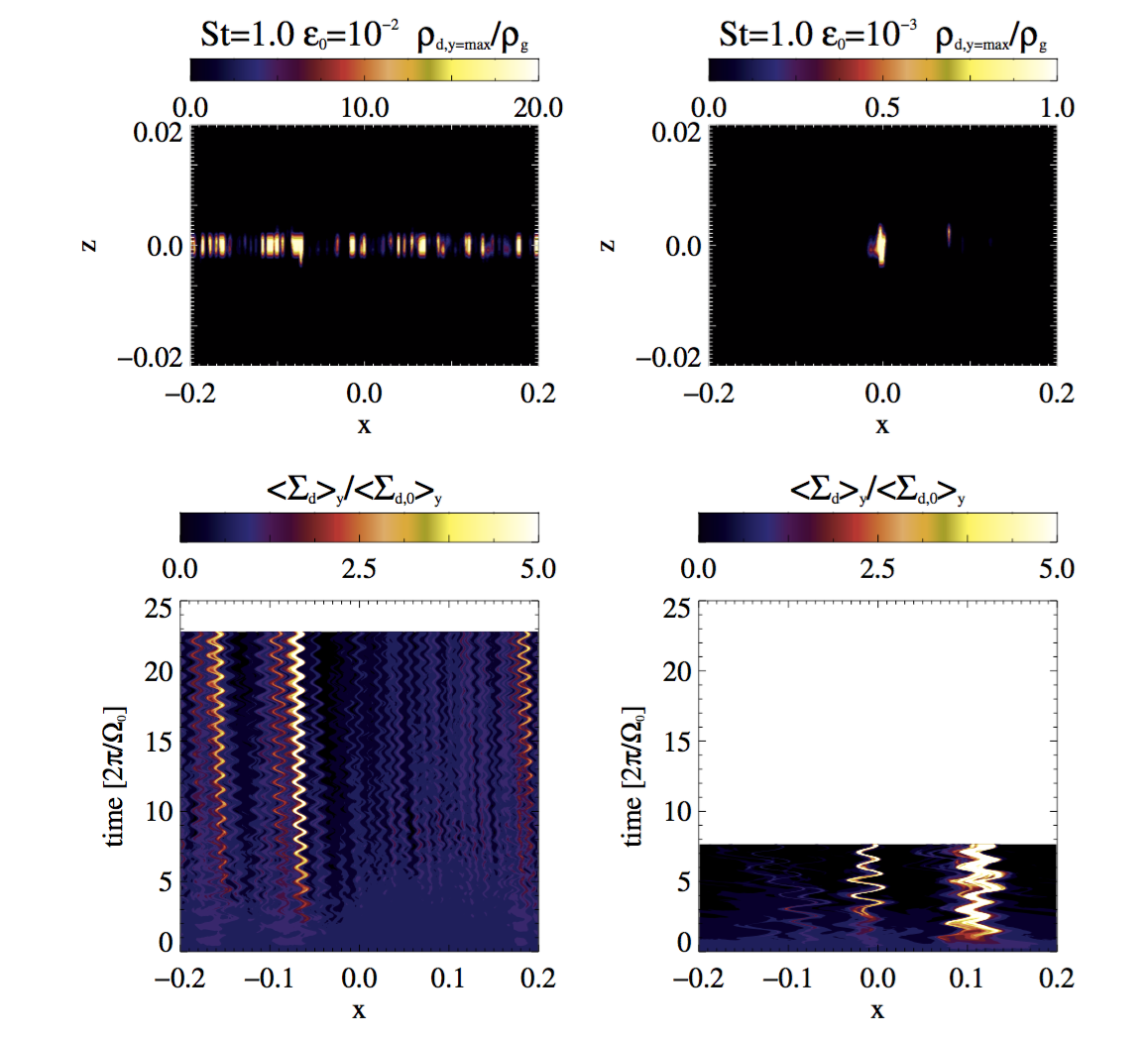}}
\end{center}
\caption{
The top row shows the dust-to-gas ratio at a meridional cut through
the box at the location of the highest dust-density. The bottom row
shows the time development of the dust surface density $\Sigma_d$
relative to its initial value $\varSigma_{d0}$ and averaged over the
entire azimuthal domain. On the left side we show the simulation of
$\St = 1$ {pebbles} with $\varepsilon_0 = 10^{-2}$ and on the right
side $\St = 1$ {pebbles} with $\varepsilon_0 = 10^{-3}$. The location
of the {pebble} concentration agrees with the location of the
vortices. The {pebble} concentration migrates inwards with the vortex
(even in the case where the vortex in the midplane is disrupted). The oscillations in the {pebble} concentration correspond to epicyclic motion of the {pebbles} within the vortex.
}
\label{fig:fig12}
\end{figure*}

\begin{figure*}
  \begin{center}
    \resizebox{.8\textwidth}{!}{\includegraphics{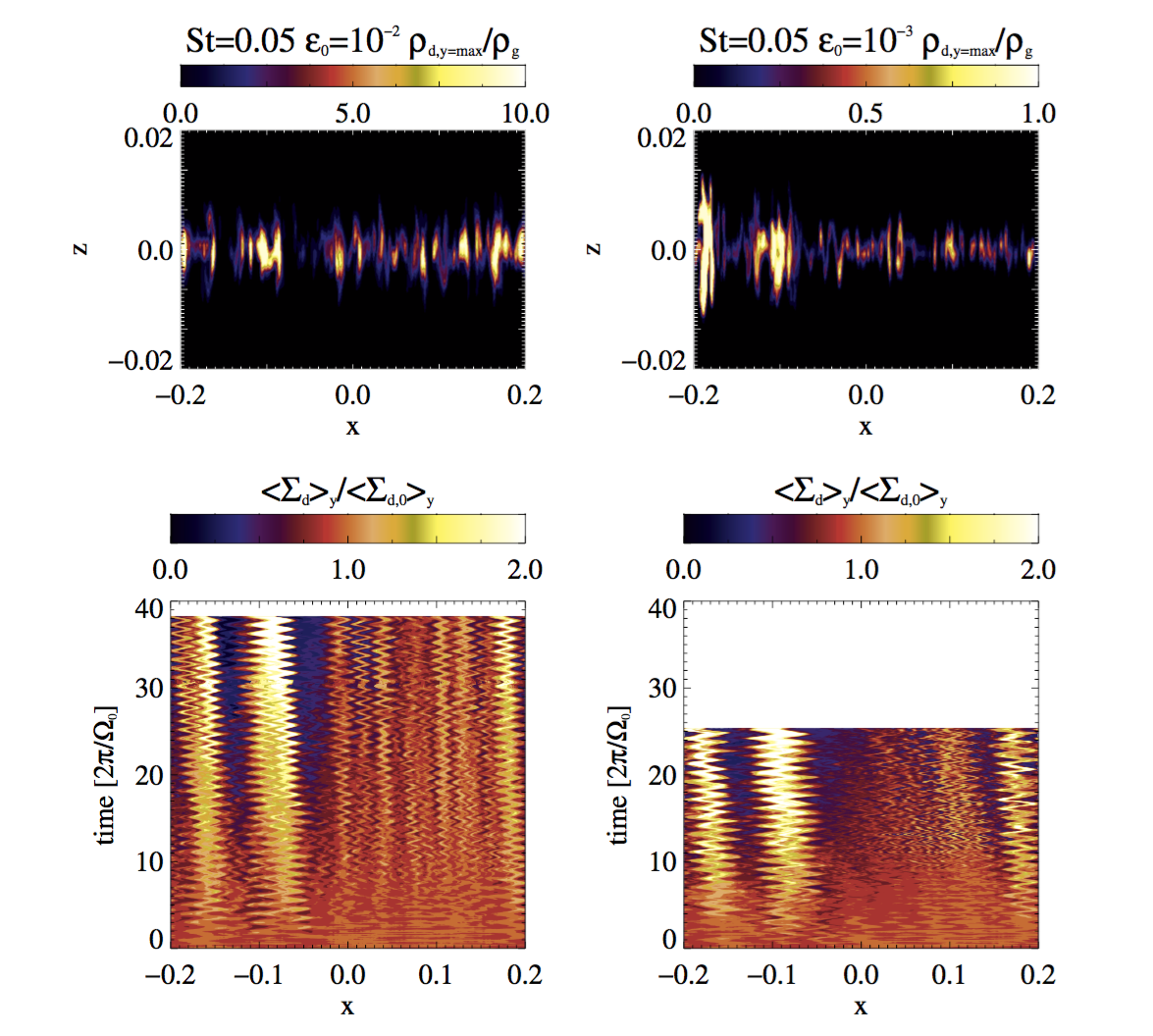}}
\end{center}
\caption{Same as \Fig{fig:fig12}, but for $\St=0.05$ pebbles. The
  vertical column of pebbles is resolved. Pebbles migrate with the
  vortex column, showing trapping is efficient even if the vortex
  column is disrupted around the midplane.}
\label{fig:fig14}
\end{figure*}

\begin{figure}
  \begin{center}
    \resizebox{\columnwidth}{!}{\includegraphics{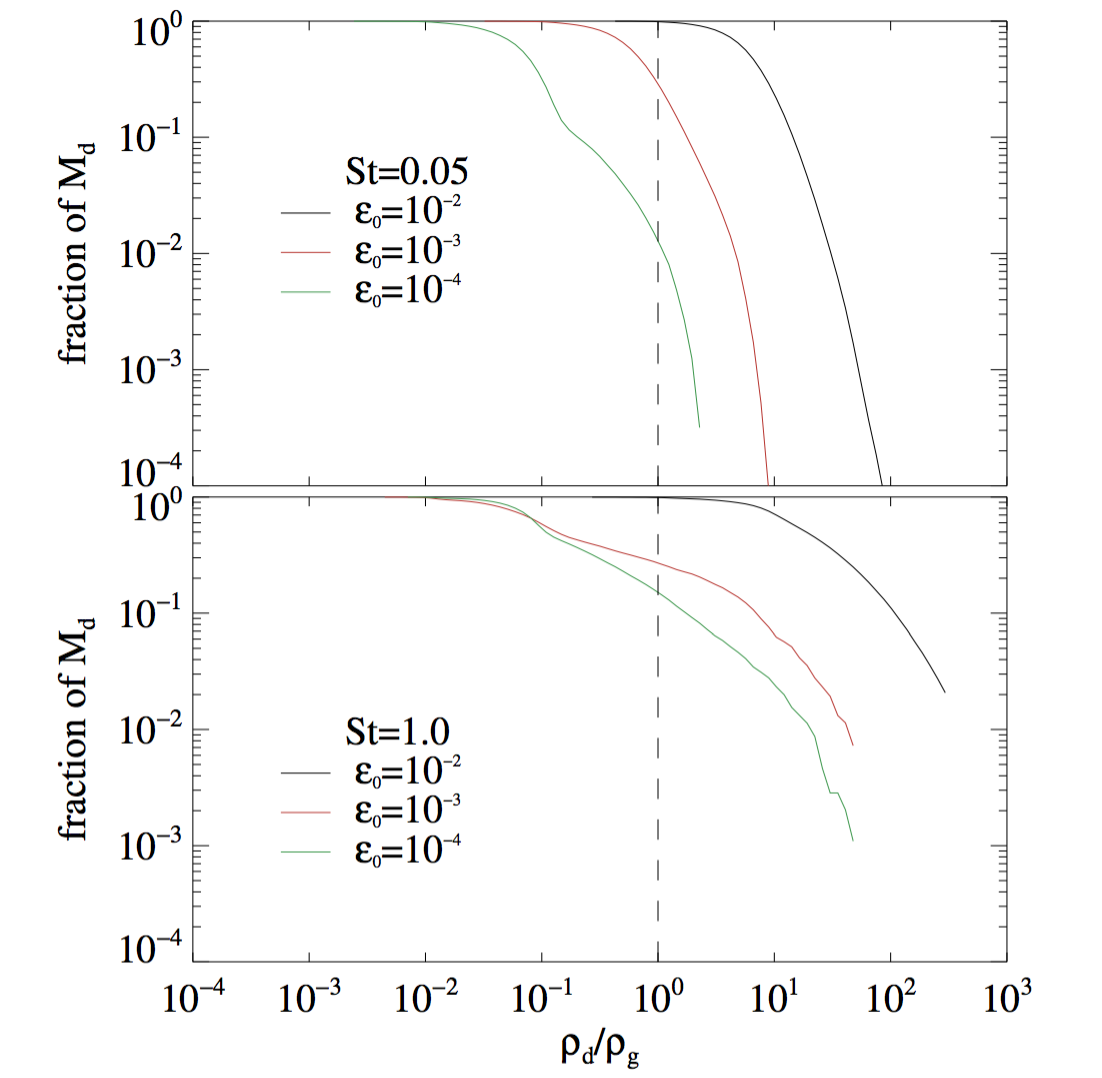}}
\end{center}
\caption{Fraction of total dust mass with dust-to-gas ratio higher than a certain value for $\St = 0.05$ and $\St = 1.0$ {pebbles} and different initial dust-to-gas ratios. For all setups the threshold for streaming instability ($\rho_d/\rho_g \geq 1$) is passed.
}
\label{fig:fig17}
\end{figure}

\subsection{Pebble evolution}
\label{sect:particlesimulations}

We first discuss the run without baroclinic driving. As
expected{,} the {pebbles} settle to the midplane, soon
triggering the streaming instability. Although we assumed an overall
initial dust-to-gas ratio of $\varepsilon_0 = 0.01$, the initial dust-to-gas ratio
in the midplane is already $\varepsilon = 1$ and increases even further as the
sedimentation progresses. The maximum dust-to-gas ratio of the simulation with $\St = 0.05$
{pebbles} and without baroclinic driving can be seen in \Fig{fig:fig7} (red
line). The other two lines in the plot show the maximum dust-to-gas
ratios for the simulation with $\xi = 2$, $\St = 0.05$ (green line) and
$\St = 1$ {pebbles} (black line).

The maximum concentration for $\St = 0.05$ with baroclinic driving is only a factor
of two higher than in the case without it. Inspecting the evolution of the dust density in the baroclinic simulation, we
see that, as expected, the {pebbles} settle to the midplane and get
trapped in the vortices. Yet, after only 3 orbits{,} the {pebbles} start
to perturb the vortices in the midplane via the streaming
instability. Like in the 2D simulations of \cite{Raettig+15}, high
{pebble} densities generate local perturbations of the velocity field
visible as steep vorticity gradients. \Fig{fig:fig8} shows the
vorticity at the top of the box (left panel), in the midplane (middle
panel) and the dust-to-gas ratio in the midplane for $\St = 0.05$. We
see that although the vortex in the midplane is heavily disrupted, the vortex column above and below the midplane is still present. Since there are no {pebbles} in the upper and lower areas of the box,
there is no disruption, and we expect these vortex columns to stay
stable over long times.

As we will discuss later, these vortex columns still have a minor influence on the {pebbles}.
The explanation for why we do not see a high increase of dust-to-gas
ratio for simulations with baroclinic vortices compared to
simulations without baroclinic driving lies in the disruption of
vortices in the midplane. To test whether there is a strong dust
concentration when vortices are stable, and also to account for the
fact that not all {pebbles} will be of the same size we perform
simulations with lower initial dust-to-gas ratios, $\varepsilon_0 = 10^{-3}$ and $\varepsilon_0 =
10^{-4}$.

Already for $\varepsilon_0 = 10^{-3}$ we see a different picture. The {pebbles} settle
to the midplane, but the vortices there are no longer disrupted: {\it the
entire vortex column is a stable feature}. {Pebbles} accumulate inside
of the vortices, as we saw in the 2D simulations, and then migrate
slowly with them. \Fig{fig:fig9} shows the midplane vertical vorticity (left panel) and
dust-to-gas ratio (right panel) for $\St=0.05$ and $\varepsilon_0 =
10^{-4}$ after 15 orbits. Strong dust concentration is achieved in
the vortex, yet no strong vortex disruption in the midplane is
observed. \Fig{fig:fig10} shows a vertical cut of the vertical
vorticity (left panels) and dust-to-gas ratio (right panel), again at
15 orbits, for the $\varepsilon_0 = 10^{-2}$ simulation (upper panel)
and the $\varepsilon_0 = 10^{-4}$ simulation (lower panel). For $\varepsilon_0 = 10^{-2}$ the vortex located around $x \approx -0.1H$ is strongly perturbed around the midplane. For $\varepsilon_0=10^{-4}$ the vortex column remains roughly undisturbed.

Table~\ref{table:table1} and \Fig{fig:fig11} show the maximum values of $\varepsilon$. We see
that in all simulations $\varepsilon > 1$ is reached and that streaming is
triggered. The baroclinic vortices prove to be an efficient {pebble}
trap, {\it even at extreme subsolar metallicity}. Even an initial
dust-to-gas ratio of $10^{-4}$ is sufficient to trigger the streaming instability inside a vortex.

\Fig{fig:fig12} shows the azimuthally-averaged pebble column density
as a function of time (bottom panels), as well as the pebble density
in the meridional plane of the maximum pebble density (upper
panels). The left hand side panels are for $\varepsilon=10^{-2}$ and the
right hand side panels for $\varepsilon=10^{-3}$, both for $\St=1$
pebbles. The same graphs but for $\St=0.05$ pebbles are shown in
\Fig{fig:fig14}. The time evolution
shows bands of high pebble concentration that are slowly migrating
radially inwards. The migration speed is equal {for} both $\St=1$ and
$\St=0.05$ {pebbles}, which evidences that this {drift} is not the pebble
flux. Indeed, this radial migration corresponds to the migration of
the vortices \citep{Paardekooper+10}. The
oscillations in these bands are the epicyclic motion of the
trapped pebbles, induced by the internal vortex motion. We
{highlight} that these are simulations where the vortex motion was
disrupted around the midplane. The vortex column, above and below the
disrupted midplane, keeps the pebbles trapped. 

We measured the intensity of pebble accumulation at different
dust-to-gas ratios. The result is shown in \Fig{fig:fig17} for the two
{pebble} sizes we consider, and three different
$\varepsilon_0$. {This quantity, the mass fraction that participates in the
streaming, is also shown in the last column of
Table~\ref{table:table1} as $\zeta$}. As
expected, simulations with higher $\varepsilon_0$ also have more mass
in large $\varepsilon$ bins. Promising for planet formation, even
simulations with $\varepsilon_0 =10^{-4}$ pass the threshold of
$\varepsilon \geq 1$, where the streaming instability sets in. While
only 1\% of $\St = 0.05$ {pebbles} accumulate in regions with
$\varepsilon \geq 1$ for $\varepsilon_0 =10^{-4}${,} this is an important
step towards higher concentrations and may eventually lead to
gravitational collapse.

\subsection{Possible Gravitational Collapse}
\label{sect:gravitationalcollapse}

In 2D runs \citep{Raettig+15} the pebble densities reached were lower
than the Roche density. But in 3D, sedimentation brings the density
above this threshold. We compare the achieved pebble
densities with the Roche density to assess if gravitational collapse
would happen had we included self-gravity in the simulation. We consider the same two disk
models as in \cite{Raettig+15}, one with the MMSN with $\Sigma_g$(5.2 AU) = 150 g\,cm$^{-{2}}$
(cases 1 and 2) and a more massive disk model with $\Sigma_g$(5.2 AU) = 800 g
cm$^{-{2}}$ (cases 3 and 4). The Roche density at the two radii that we
consider is $\rho_R(1 AU) = \xtimes{2.83}{-7} \ {\rm g\,cm}^{-3}$ (case 1 and 3) and $\rho_R(5.2
AU) = \xtimes{2.01}{-9} \ {\rm g\,cm}^{-3}$ (case 2 and 4).

In Table~\ref{table:table2} we show the maximum dust-to-gas ratios
that were reached at the end of the computational time. The last four
columns show the ratio of $\rho_R$ and the gas density for the two
disk profiles and the two radii. If $\varepsilon_{\rm max}$ exceeds
this ratio, it means that the local dust-to-gas ratio exceeds $\rho_R$
at that radius, and the pebble accumulation would collapse in the
absence of diffusion (indicated by bold face in
Table~\ref{table:table2}).

We see that at 1 AU the maximum dust densities are lower than the
Roche density for all runs except for $\St = 1$ {pebbles} with
$\varepsilon_0 = 0.01$. Yet if we go to 5.2AU there are three more
runs for the more massive disk model that exceed the critical
$\rho_R$. This means that the clumps with these high overdensities
should collapse and form bodies that are held together by their own
gravity. It is easier to form planetesimals at larger radii, because
$\rho_R$ falls off {more steeply} than $\rho_g$. Therefore, the
ratio $\rho_R/\rho_g$ decreases radially, making it easier for dust
densities to {cross} the critical density {threshold}.

\begin{table*}
\caption{Dust density compared to Roche density.}
\label{table:table2}
\begin{center}
\begin{tabular}{c c c c c c c}\hline
Run            &$\varepsilon_{\rm max}$& $\rho_R/\rho_g$ & $\rho_R/\rho_g$ & $\rho_R/\rho_g$ & $\rho_R/\rho_g$ \\\hline
                  && case 1 (5.2AU) & case 2 (1AU) & case 3 (5.2AU) & case 4 (1AU) \\\hline
{\bf 3DF05}       &111.2&262 & 1607 &{\bf 49} &301\\ 
3DF05E-3  &8.7& 262 & 1607 &49 &301\\
3DF05E-4  &2.1& 262 & 1607 &49 &301\\
{\bf 3DF1}         &945.4& {\bf 262} & 1607 &{\bf 49} &{\bf 301}\\
{\bf 3DF1E-3}   &227.7& 262 & 1607 &{\bf 49} &301\\
3DF1E-4   &34.2& 262 & 1607 &49 & 301\\\hline
\end{tabular}
\end{center}
\end{table*}

\subsection{{Limitations}}

{
One limitation of the work is that we include vertical gravity only
for the pebbles, treating the gas as unstratified. We can assess how
important this would be. Since the pebble
scale height $H_d$ is much smaller than the gas scale height $H$, the
gas density is essentially constant over a distance $H_d$. At one dust
scale height, the gas density is 99.0\% and 99.95\%  of the original
density for $\St=0.05$ and $\St=1$, respectively. We do not expect this
variation to be dynamically significant. An important point,
though, is the exclusion of vertical shear, given by \citep{LyraUmurhan19}
\beq
\frac{d\varOmega}{dz} =\frac{1}{2} \varOmega  h^2 q_T \frac{z}{H^2} 
\eeq
This vertical shear can in principle be of dynamical consequence (even
though the oscillations leading to vertical shear instability are
stabilized by buoyancy in our setup with finite cooling time). Yet,
let us compare this gravity-induced vertical shear to the shear the
pebbles impart on the gas, due to the backreaction of
the drag force. In the absence of gas, the pebbles would move at the
Keplerian rate. Where the pebbles dominate dynamically (at dust-to-gas
ratios above unity), in the midplane, the drag force backreaction
accelerates the gas layer to move at the Keplerian rate. This 
establishes a strong localized vertical shear in the gas that, in
turn, breaks into Kelvin-Helmholtz instability \citep{Johansen+06} generating turbulence
and diffusion. This velocity difference is of magnitude $\eta u_k$, so
the shear {over a distance $z$} is 
\beq
\frac{d\varOmega}{dz} = \frac{1}{2}\varOmega  h^2 \ksi {z^{-1}}.
\eeq
So, for $\xi = q_T$ (ignoring the density gradient for the argument), the ratio of the shear given by gravity and the
one caused by the pebbles is 
\beq
{f \equiv} \frac{\left(d\varOmega/dz\right)_{\rm gravity}}{\left(d\varOmega/dz\right)_{\rm pebbles}} = \frac{z^2}{H^2}
\eeq
\noindent which is vanishingly small in the midplane. For $z=H_d$ it
yields $f = H_d^2/H^2 = \delta/(\St+\delta)$. For $\delta\approx 10^{-3}$,
this translates into $f \approx 10^{-2}$ for $\St=0.05$ and
$f \approx 10^{-3}$ for $\St=1$. That is, the pebbles
dominate the vertical shear as long as $\delta \ll \St$.
Fundamentally, the approximation is justified because the pebble layer
is very thin in comparison to the gas layer. Still, the effect of
vertical gravity on the gas, even if small, should be considered in future work.}

\section{Conclusion}
\label{sect:conclusion}

In this paper we have performed three-dimensional simulations of
self-sustained baroclinic vortices, generated by the convective
overstability, along with Lagrangian particles to simulate $\St = 0.05$ and $\St = 1$ pebbles
in the disk, including the drag force and its
backreaction. {These values of $\St$ correspond roughly to
  cm-sized pebbles in the inner disk ($\sim$ 5\,AU) and mm-sized pebbles in the outer
  disk ($\sim$ 50\,AU) for $\St=0.05$, and m-sized boulders in the inner disk and
  cm-sized pebbles in the outer disk for $\St=1.$} While the gas is
unstratified, we included vertical gravity for the pebbles, to allow
for sedimentation. {\cite{Fu+14} report, via long evolution
  time 2D simulations of Rossby vortices, a decreased vortex
  lifetime as the dust feedback increases, either as a result of
  higher dust-to-gas ratio of larger pebble sizes. In our 3D
  simulations, and consistent with these 2D results}, for $\varepsilon_0 =10^{-3}$ or $\varepsilon_0 =
  10^{-4}$ the vortex midplane is not disrupted, remaining 
  coherent over the simulation period.  {Also consistent with the 2D results}, for initial dust-to-gas ratio of $\varepsilon_0 =
10^{-2}$ we see that the vortex column in the midplane is disrupted
after a very short time of pebble concentration. {\cite{Raettig+15} report} a similar
picture in the two-dimensional simulations for $\St = 1$ {pebbles} when
the local dust-to-gas ratio exceeded a value of $\approx$ 10. Due to
the sedimentation of dust, this dust-to-gas ratio is reached after
only a few orbital periods for both {pebble} sizes.

Yet, {here we show with 3D models that} because the {pebble} sedimentation occurs only in a thin layer around the
midplane, the vortex column remains coherent above and below the
midplane, and retains its ability to trap pebbles. The pebbles
disrupt the vortex {flow within the thickness of the pebble layer}, but they do not destroy {the
vortex column}. The initial
concentration in the vortex grows as the simulations progress. The
pebble overdensity drifts not at the pebble flux rate, but rather in
the slower migration timescale of the vortex, evidencing efficient
trapping. {Our} result{s are} important because based on the previous 2D result suggesting {eventual}
disruption, the vortex interpretation of ALMA observations has been called into question. We show instead that the
vortex behaves like a Taylor column, and the pebbles as obstacles to the flow.

{For lower dust-to-gas ratio}, although the concentration remains below the high concentrations seen
for $\varepsilon_0 = 0.01${,} they still reach local enhancements of $\varepsilon > 1$ and{, consequently,} trigger the streaming instability.
To determine whether the local {pebble} density enhancements can cause
gravitational collapse were we to turn on {pebble} self-gravity, we
compare their density with the Roche density. Depending on {the} disk model
and radial position in the disk{,} we see that $\St = 0.05$ pebbles with
a high initial dust-to-gas ratio, or $\St = 1$ pebbles with
$\varepsilon_0 = 10^{-{3}}$ can exceed the Roche density. Had we
considered selfgravity, they would have formed gravitationally bound
objects.

We conclude that baroclinic vortices in 3D are an efficient mechanism
for trapping and concentrating {pebbles}, allowing streaming
instability and planetesimal formation even at subsolar
metallicity. In a future work we will include selfgravity in order to
assess the mass distribution formed in these baroclinic vortices.

\acknowledgments

We acknowledge conversations with Nienke van der Marel.
W. L. acknowledges support from NSF through grant AST-2007422, and from NASA
through grant 20-TCAN20-0011.
The simulations presented in this paper utilized the Stampede
cluster of the Texas Advanced Computing Center (TACC) at The
University of Texas at Austin, through XSEDE grant TG-AST140014.
{This collaboration was made possible through the support of
the Annette Kade Graduate Student Fellowship Program at
the American Museum of Natural History.}

\end{document}